\documentclass[a4paper]{spie}  

\usepackage{amsmath,amsfonts,amssymb}
\usepackage{graphicx}
\usepackage[colorlinks=true, allcolors=blue]{hyperref}
\usepackage{caption}
\usepackage{subcaption}
\usepackage{makecell}
\usepackage{color}
\usepackage{colortbl}
\usepackage{changepage}

\title{Efficient subtyping of ovarian cancer histopathology whole slide images using active sampling in multiple instance learning}

\author[a]{Jack Breen}
\author[b]{Katie Allen}
\author[b]{Kieran Zucker}
\author[c]{Geoff Hall}
\author[b]{Nicolas M. Orsi*}
\author[a]{Nishant Ravikumar*}
\affil[a]{CISTIB Centre for Computational Imaging and Simulation Technologies in Biomedicine, School of Computing, University of Leeds, Leeds, United Kingdom}
\affil[b]{Leeds Institute of Medical Research (LIMR), University of Leeds, Leeds, United Kingdom}
\affil[c]{Leeds Teaching Hospitals NHS Trust (LTHT), Leeds, United Kingdom}

\authorinfo{Corresponding author: Jack Breen, scjjb@leeds.ac.uk.     \\
*Indicates joint last authors.
}

\begin{document} 
\maketitle

\begin{abstract}
Weakly-supervised classification of histopathology slides is a computationally intensive task, with a typical whole slide image (WSI) containing billions of pixels to process. We propose Discriminative Region Active Sampling for Multiple Instance Learning (DRAS-MIL), a computationally efficient slide classification method using attention scores to focus sampling on highly discriminative regions. We apply this to the diagnosis of ovarian cancer histological subtypes, which is an essential part of the patient care pathway as different subtypes have different genetic and molecular profiles, treatment options, and patient outcomes. We use a dataset of 714 WSIs acquired from 147 epithelial ovarian cancer patients at Leeds Teaching Hospitals NHS Trust to distinguish the most common subtype, high-grade serous carcinoma, from the other four subtypes (low-grade serous, endometrioid, clear cell, and mucinous carcinomas) combined.  We demonstrate that DRAS-MIL can achieve similar classification performance to exhaustive slide analysis, with a 3-fold cross-validated AUC of 0.8679 compared to 0.8781 with standard attention-based MIL classification. Our approach uses at most 18\% as much memory as the standard approach, while taking 33\% of the time when evaluating on a GPU and only 14\% on a CPU alone. Reducing prediction time and memory requirements may benefit clinical deployment and the democratisation of AI, reducing the extent to which computational hardware limits end-user adoption.  
\end{abstract}

\keywords{Digital pathology, ovarian carcinoma, computer vision, histology, computer-aided diagnosis}

\section{BACKGROUND}
\begin{figure}[h]
\centering
\includegraphics[width=\textwidth]{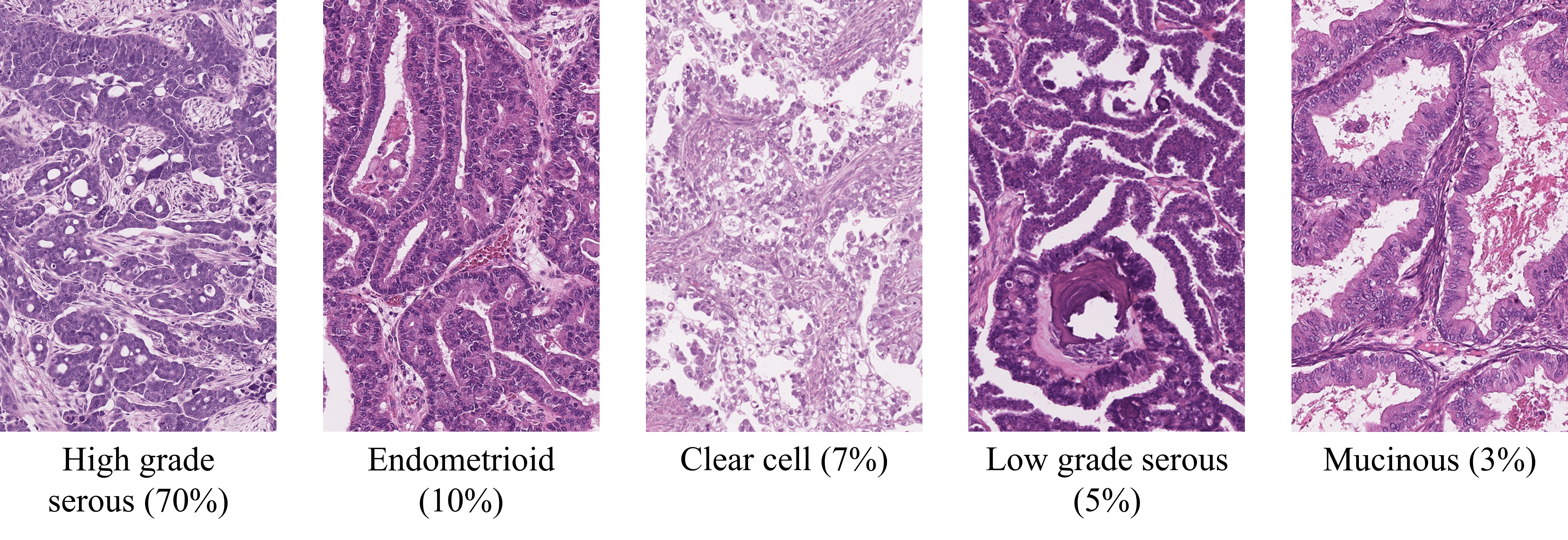}
\caption{Examples of the five major morphological subtypes of epithelial ovarian cancer, with corresponding frequencies \cite{Kossai2018}. These frequencies do not sum to 100\% due to the existence of rarer subtypes which are not shown.}
\label{fig:subtypes}
\end{figure}

Ovarian cancer is the eighth most common cancer in women, with over 300,000 new cases diagnosed worldwide each year\cite{Sung2021}. The disease carries a significant mortality burden, accounting for 200,000 deaths per year \cite{Sung2021}. There are five major morphological subtypes of epithelial ovarian cancer, examples of which are shown in Figure \ref{fig:subtypes}. The different morphologies represent significantly different diseases, with distinct molecular features, prognoses, and responses to treatment \cite{Kossai2018}. Pathologists diagnose these through rigorous and time-consuming microscopic analysis of pathology slides. Determination of the carcinoma subtype can be challenging for general pathologists on haematoxylin and eosin (H\&E)-stained tissue slides, often requiring the use of immunohistochemistry (IHC) or referral to specialist gynaecological pathologists. The increasing availability of digital pathology scanners has provided opportunities for the creation of computer-aided diagnostic tools which may help to improve the speed, accuracy and objectivity of diagnosis. It is hoped that these tools might help to support chronically understaffed pathology departments, such as those in the UK's National Health Service (NHS), most of which outsource work or hire temporary locums to meet demand \cite{RCPath2018}. This is a global issue, with the vast majority of countries having even fewer pathologists per capita than the UK \cite{wilson2018}. 

Multiple instance learning (MIL) is a supervised learning approach in which multiple \emph{instances} with shared properties are aggregated into a \emph{bag} for modelling. A histopathology whole slide image (WSI) can be represented as a bag, with the instances being smaller patches selected from the image. Attention-based multiple instance learning (ABMIL) \cite{Ilse2018} is a state-of-the-art approach for weakly supervised WSI classification in which each patch is assigned an attention score, which is then used to weight the importance of the patch when aggregating patch encodings to generate a WSI encoding. This is a computationally intensive process due to the gigapixel WSIs each being split into tens of thousands of patches, each of which is separately encoded before aggregation through a weighted average. The WSI encodings are ultimately used for slide-level classification through a neural network.

\subsection{Related Work}
\subsubsection{Within-bag instance sampling}
When using MIL in a setting with many instances per bag, such as modelling a WSI as a bag of many patches, it may be pertinent to use within-bag instance sampling to effectively reduce the number of instances. This may be used to focus on relevant instances, improve robustness to outlier instances, or to reduce the overall computational burden by not fully processing all instances. Within-bag sampling can be as simple as randomly selecting instances\cite{Lerousseau2021, Zhu2017}, or only selecting instances within a specific region when using spatially related instances, such as patches from within an image\cite{Couture2018}. Multiple magnifications of a histopathology slide can be efficiently leveraged by random instance selection across magnifications \cite{BenTaieb2017}, or by performing discriminative region detection on the lower-magnification (smaller) image to guide instance selection on the higher-magnification (larger) image \cite{Katharopoulos2019}. 

Within-bag sampling has previously been integrated with ABMIL by splitting each bag into a group of \emph{mini-bags} - overlapping subsets of the original bag \cite{Koriakina2021}. Each mini-bag is classified through an ABMIL model which has been trained on mini-bags, and the bag classifications are given by the majority vote of the mini-bags. Mini-bag processing reduces memory requirements, but the duplication of instances across multiple mini-bags is likely to increase inference time. Further, as the key instance detection is based upon the ABMIL attention weights, all instances need to be passed through the feature extraction part of the network, which has a high computational burden. Subsequent work \cite{Koriakina2022} showed this approach to be less accurate than conventional single instance learning for cytological data, but it has not been evaluated for whole slide histopathological data, where single instance learning is not practical due to the large image sizes.

Some MIL sampling approaches use patch classification probabilities rather than attention scores. One such approach is top-k sampling \cite{Campanella2019}, where all patches are evaluated and those with the highest patch classification scores are used for slide-level classification. Monte-Carlo sampling \cite{Combalia2018} instead uses a random sample initially, then iteratively replaces the patches with the lowest individual classification scores with new random patches to improve the discriminative ability of the sample. Patches can also be sampled using expectation maximisation\cite{Hou2016}. It is not clear that any of these classification approaches can offer efficiency improvements, with top-k sampling and expectation maximisation requiring all patches to be processed through a neural network before sampling, and Monte-Carlo sampling reported to be slower than whole slide processing. While these classification approaches have not demonstrated an increase in efficiency, similar approaches have been shown to benefit WSI segmentation speed without sacrificing accuracy\cite{Broad2022,CruzRoa2018}.



\subsubsection{Ovarian cancer subtyping}
Ovarian cancer subtyping has included multi-resolution feature extraction\cite{BenTaieb2015,BenTaieb2016,BenTaieb2017}, mitigation of domain differences through normalisation\cite{Boschman2022} and data synthesis\cite{Levine2020}, and tumour segmentation before classification\cite{Farahani2022}. Previous studies focused almost exclusively on maximising classification accuracy and AUC scores, where we aim to balance this predictive performance with computational efficiency. Dataset sizes have frequently been a limiting factor of ovarian cancer subtyping research, with older studies having fewer than 100 total slides across the 5 subtypes\cite{BenTaieb2015,BenTaieb2016}, and more recent research having fewer than 400. Only one study has used a larger ovarian dataset for subtyping, with 1008 WSIs from 545 patients\cite{Farahani2022}, bringing it in line with cutting-edge research in other malignancies, such as renal, lung, and breast cancers\cite{Lu2021}. The method proposed to utilise the 1008 ovarian cancer WSIs was a two-stage approach to improve evaluation efficiency, with patches only encoded for classification if they were determined to contain tumour tissue by a segmentation model\cite{Farahani2022}.

\section{METHODS}
\subsection{Data}
Our dataset comprises 714 histopathology slides from 147 patients with epithelial ovarian cancers who had tissue resected during staging or interval debulking surgery at Leeds Teaching Hospitals NHS Trust between 2008 and 2021. Subjects were selected by a pathologist to create a dataset of the five major subtypes, in similar proportions to those seen in the general population. We required that tumours were carcinomas of tubo-ovarian-primary peritoneal origin, that a single epithelial subtype was present, and that associated clinical metadata was available in the patient health records. For each of the subjects, the resection slides were reviewed by at least one of a team of two pathologists, and their unbiased diagnostic opinion was compared to the originally reported diagnosis by a specialist gynaecological pathologist. Any cases with diagnostic discrepancies were excluded. A representative subset of diagnostic tumour slides of formalin-fixed, paraffin-embedded (FFPE) H\&E-stained tissue was selected from each case, most of which contained adnexal tissue, though other sites of ovarian cancer were included.   
The slides had any mounting artifacts corrected, and were cleaned and anonymised. They were digitised on a Leica AT2 scanner at 40x magnification. Overall, 455 slides were collected from 92 patients diagnosed with high-grade serous carcinoma (HGSC), 75 slides from 14 patients with low-grade serous carcinoma (LGSC), 60 slides from 15 patients with clear cell carcinoma (CCC), 76 slides from 15 patients with endometrioid carcinoma (EC), and 48 slides from 11 patients with mucinous carcinoma (MC). 

\subsection{Modelling}
\subsubsection{Baseline classifier}
\begin{figure}[h]
\centering
\includegraphics[width=\textwidth]{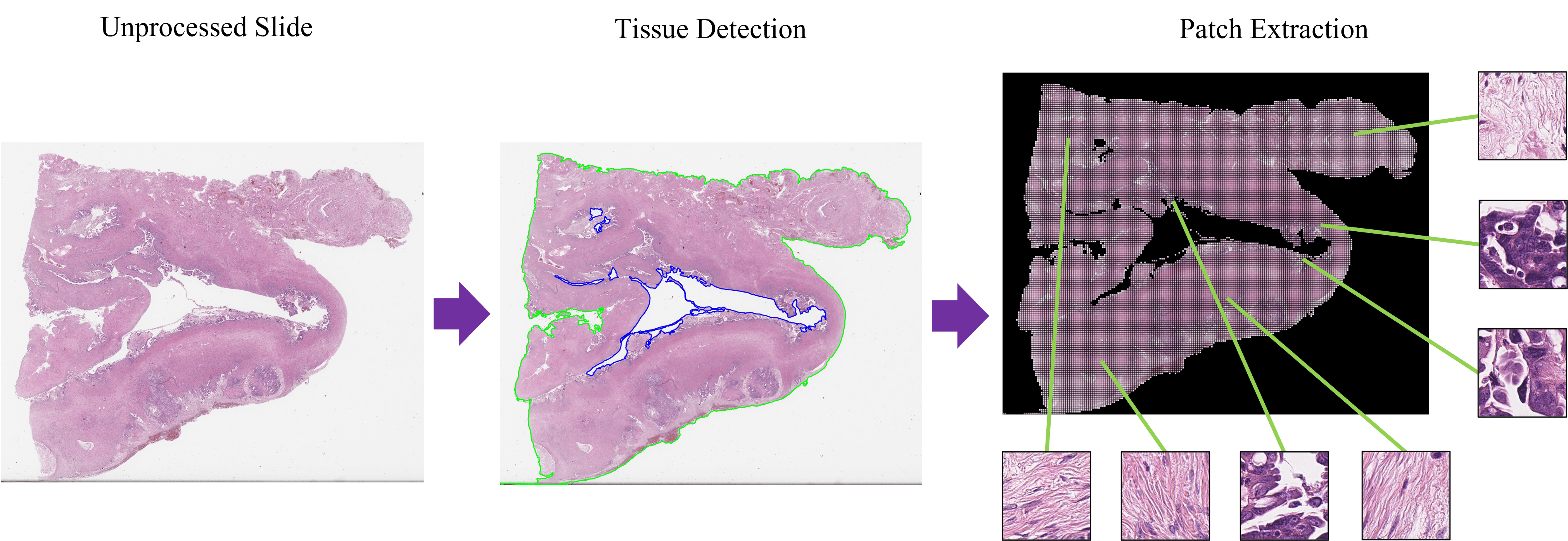}
\caption{The patch extraction process for performing multiple instance learning with a digital pathology slide, where each patch is an \emph{instance} and the entire slide is a \emph{bag}. Here patches are 256x256 pixels from a 20x magnification pathology slide. Tissue detection is performed using saturation thresholding.}
\label{fig:patching}
\end{figure}
We performed a binary classification of ovarian cancer WSIs, classifying HGSC against all other subtypes. Approximately 70\% of all epithelial ovarian cancer cases are HGSCs \cite{Kossai2018}, so this classifier aimed to detect less common histological subtypes. We chose not to perform 5-class classification because, although our dataset is one of the largest datasets ever used for this task to date, we have a limited quantity of data from the least common subtypes which is insufficient for robust analysis. We use ABMIL with sigmoid gated-attention, where the attention weight of a patch is calculated using
\begin{equation}
a_k = \frac{\exp\{\mathbf{w}^\top(\tanh(\mathbf{V}\mathbf{h}_k^\top)\odot\sigma(\mathbf{U}\mathbf{h}_k^\top))\}}{\sum_{j=1}^{K}\exp\{\mathbf{w}^\top(\tanh(\mathbf{V}\mathbf{h}_j^\top)\odot\sigma(\mathbf{U}\mathbf{h}_j^\top))\}},
\end{equation}
for \emph{1}$\times M$ instance embedding $\mathbf{h}_k\in\{\mathbf{h}_1,...,\mathbf{h}_K\}$, $L\times $\emph{1} parameter vector $\mathbf{w}$, and $L\times M$ parameter matrices $\mathbf{U},\mathbf{V}$, where $\odot$ is an element-wise multiplication and dimensions $L$ and $M$ are hyperparameters.

We split each slide into 256x256 pixel non-overlapping patches at 20x magnification as shown in Figure \ref{fig:patching}. We encoded these patches using a static ResNet50 convolutional neural network pre-trained on ImageNet. Following this, we used the sigmoid gated-attention to 
generate whole slide embeddings of length 1024 and classified these through a fully connected neural network. We tuned hyperparameters through a random search, as detailed in Appendix \ref{sec:hyperparams}.

\subsubsection{DRAS-MIL}


We propose Discriminative Region Active Sampling for Multiple Instance Learning (DRAS-MIL) in this paper, a novel approach for efficient slide classification (during inference). It starts by taking a completely random sample of patches from the WSI, classifying these through the trained ABMIL model, and collecting the attention scores of processed patches. Sampling weights for subsequent samples are assigned based on these attention scores, with higher weights in spatial regions around high-attention patches. This mimics the process used by pathologists, who will not look at an entire slide at high magnification, but instead sample regions, giving more attention to the spatial area around other diagnostically relevant tissue. We include a hyperparameter \emph{sampling\_random} which allows a proportion of the samples to be drawn randomly rather than through attention-based weighted sampling. This hyperparameter controls the level of exploration of new regions and the level of exploitation of the best regions already found, with higher \emph{sampling\_random} giving more exploration and lower \emph{sampling\_random} giving more exploitation. 

We chose to use 800 patches per WSI in the sampling evaluations, which represents approximately 5\% of the tissue area in a typical slide (the slides have 740-33961 patches each, with an average of 15990 and median of 16230). Any slide with less than 800 patches was evaluated with whole slide processing, though this only applied to one slide in our entire dataset. 800 samples are enough to generate sampling weights for the majority of patches in a slide, as shown in Figure \ref{fig:sampling}. We used a random search to tune and set five sampling hyperparameters, as described in Appendix \ref{sec:hyperparams}.

We evaluated the classification performance of our sampling approach through 3-fold cross-validation, with a third of the patients assigned to each of the training, validation, and testing sets, stratified by ground truth subtype diagnosis. We compared this approach to the random sampling of 800 patches, and to the baseline MIL approach of evaluation using all available patches.
To account for randomness, we repeated each sampling approach 50 times and performed 100,000-epoch bootstrapping, where each slide was represented exactly once per epoch by one of the 50 predictions made for the slide. Repeating this 100,000 times gave 100,000 values for each metric, from which we calculated the average and standard deviation. 

We evaluated the efficiency of DRAS-MIL and the baseline model using a test set of 50 randomly selected WSIs which was consistent across experiments. This smaller test set was used because efficiency experiments were much slower to run than accuracy experiments - during the latter we pre-extracted all patch features to speed up repeated evaluations, but during efficiency experiments we extracted features in real-time to better represent the clinical workflow. We measured the average time taken to evaluate a WSI and the maximum GPU memory requirements. There is a trade-off between the time to evaluate images and memory requirements, so we measured both of these metrics for different batch sizes (1, 4, 8, 16, 32, and 64), which represent the maximum number of patches processed concurrently.  Efficiency experiments alternated between active sampling and full slide processing, running each three times and taking the median value for each batch size as the true value. All efficiency experiments were conducted on a desktop computer with a single NVIDIA GTX 1660 GPU with 6GB of VRAM, an Intel i5-4460 CPU @ 3.2GHz, and 16 GB of RAM. Model training and hyperparameter tuning were conducted on an NVIDIA DGX A100 server with 8 NVIDIA A100 GPUs and 256 AMD EPYC 7742 CPUs @ 3.4GHz.

\begin{figure}
\centering
\begin{subfigure}[b]{0.3\textwidth}
\centering
\includegraphics[width=\textwidth]{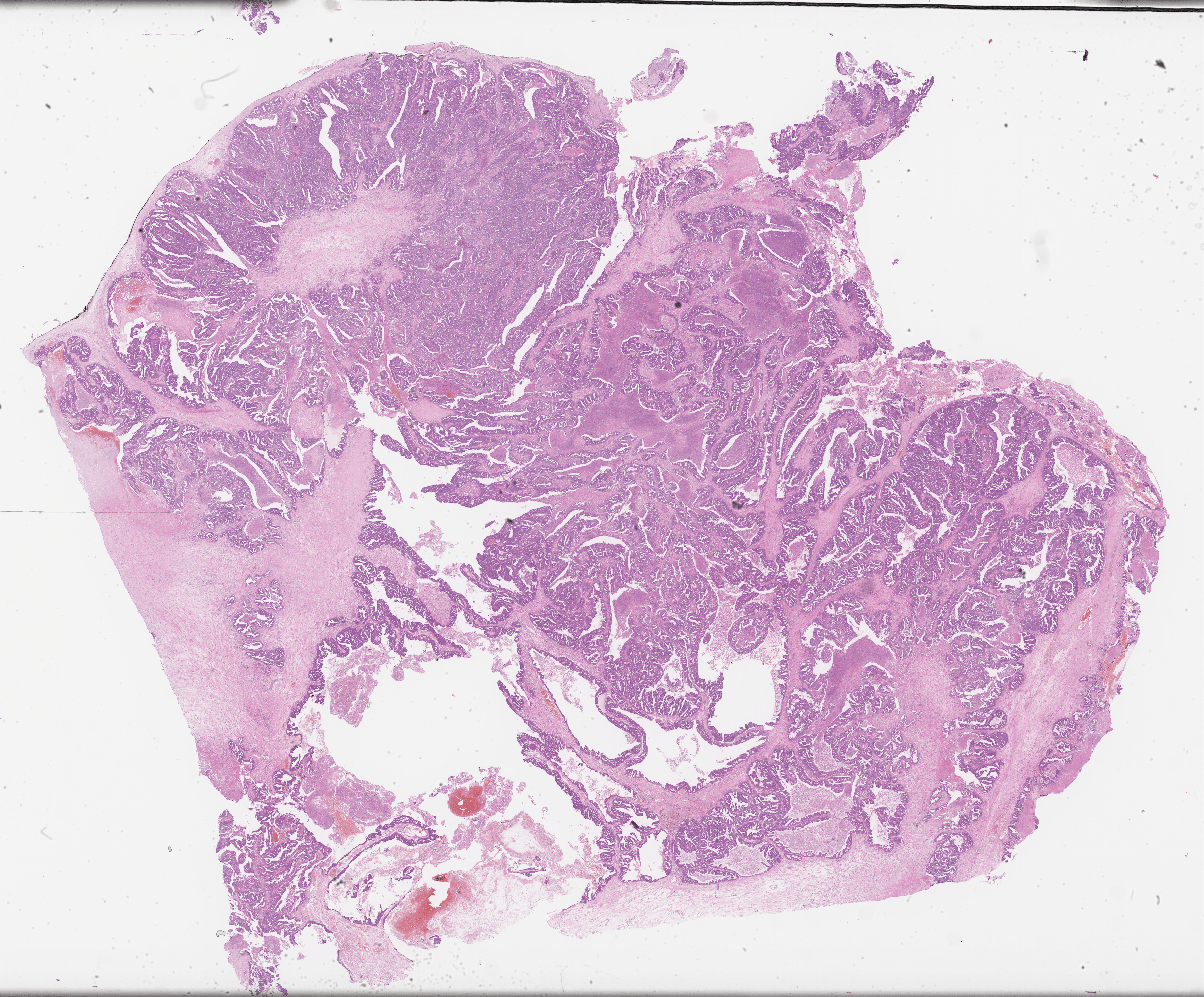}
\caption{Unprocessed whole slide image}
\label{fig:thumbnail}
\end{subfigure}
\hspace{1cm}
\begin{subfigure}[b]{0.3\textwidth}
\centering
\includegraphics[width=\textwidth]{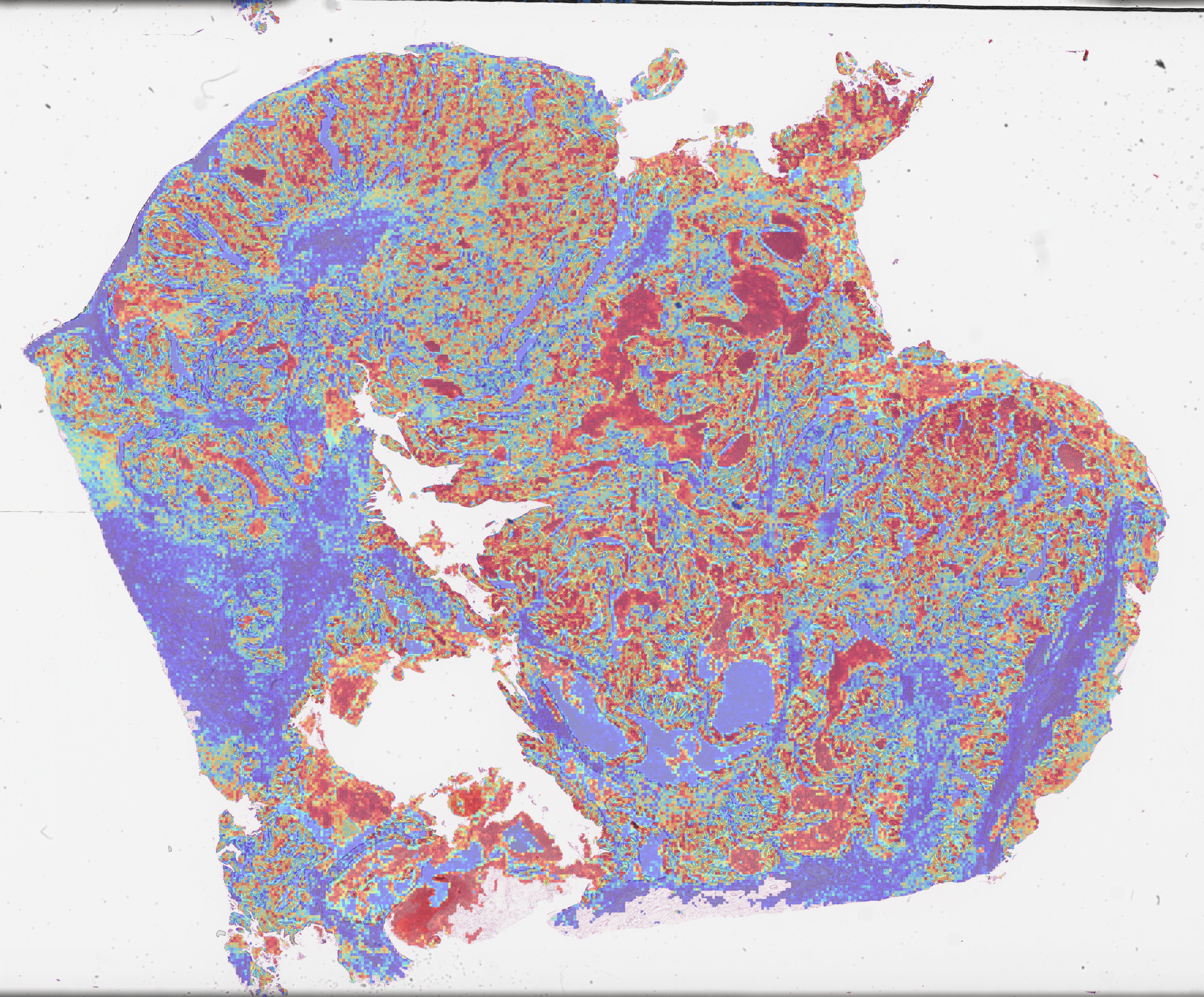}
\caption{ABMIL attention scores}
\label{fig:heatmap}
\end{subfigure}
\caption{Attention scores from attention-based multiple instance learning (ABMIL) whole slide processing}
\label{fig:heatmaps}
\end{figure}

\begin{figure}
\centering
\begin{subfigure}[b]{0.3\textwidth}
\centering
\includegraphics[trim = 0.5cm 1cm 1.5cm 0cm, width=\textwidth]{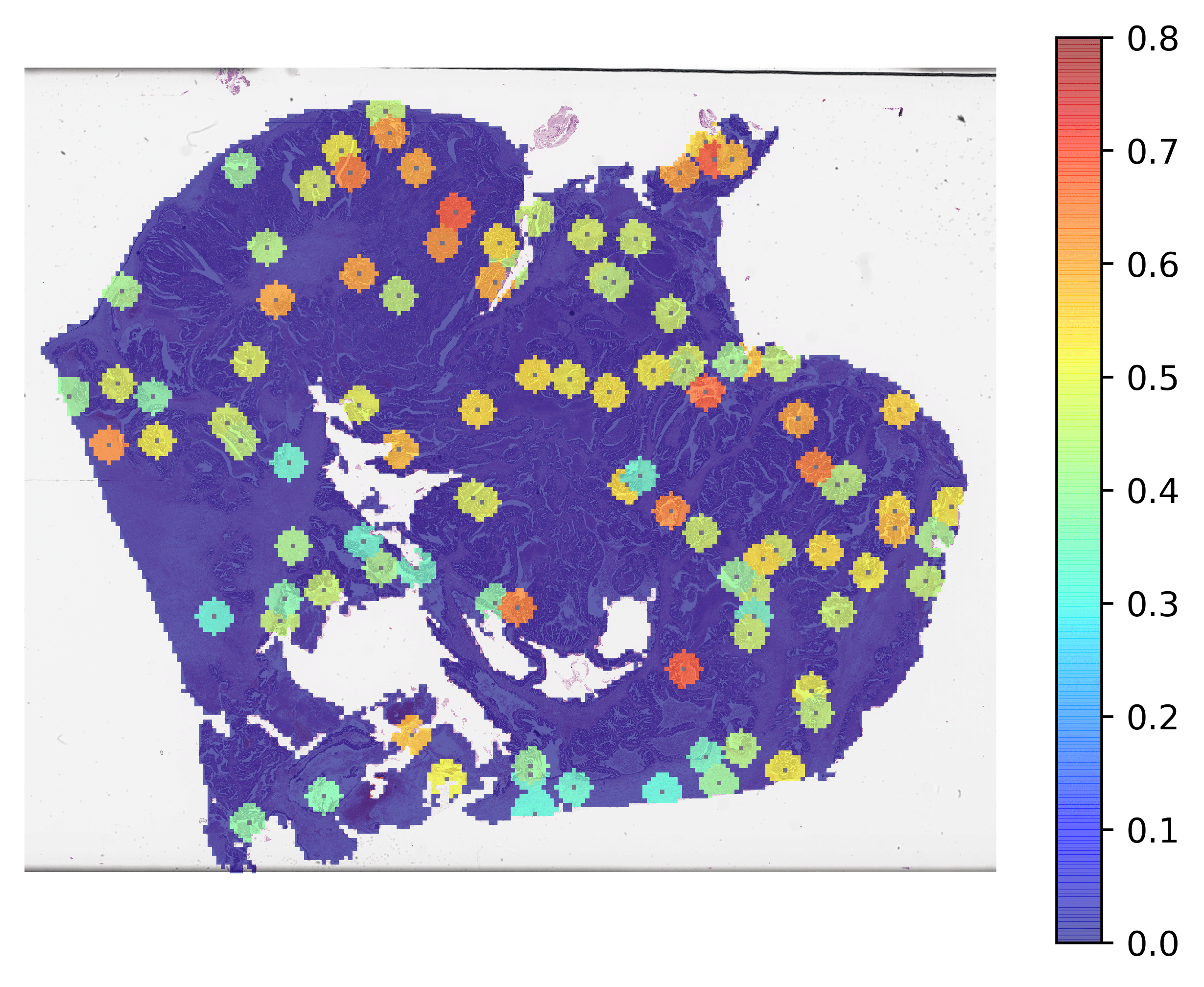}
\caption{100 Samples}
\label{fig:it1}
\end{subfigure}
\begin{subfigure}[b]{0.3\textwidth}
\centering
\includegraphics[trim = 0.5cm 1cm 1.5cm 0cm,width=\textwidth]{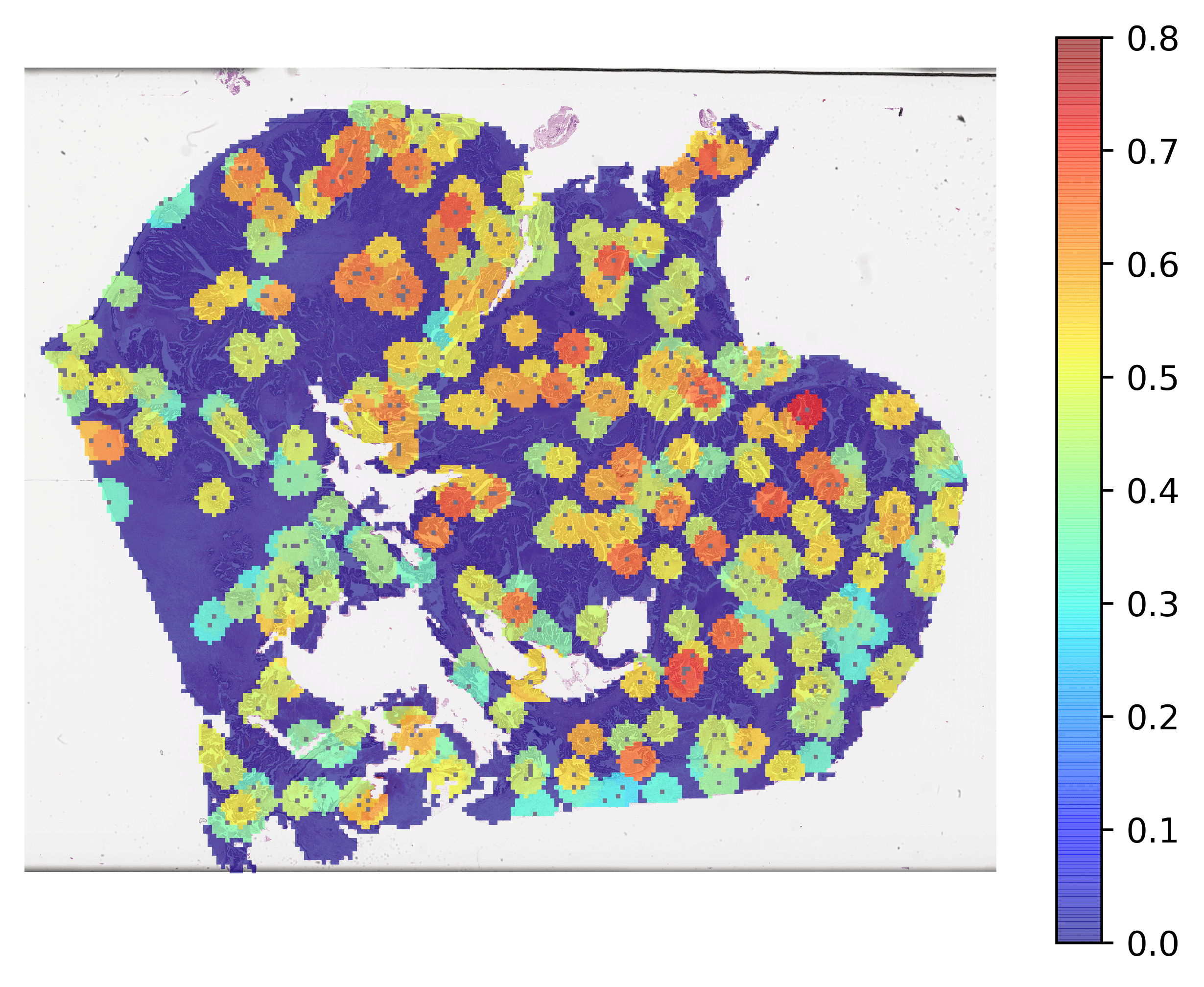}
\caption{400 Samples}
\label{fig:it6}
\end{subfigure}
\begin{subfigure}[b]{0.3\textwidth}
\centering
\includegraphics[trim = 0.5cm 1cm 1.5cm 0cm, width=\textwidth]{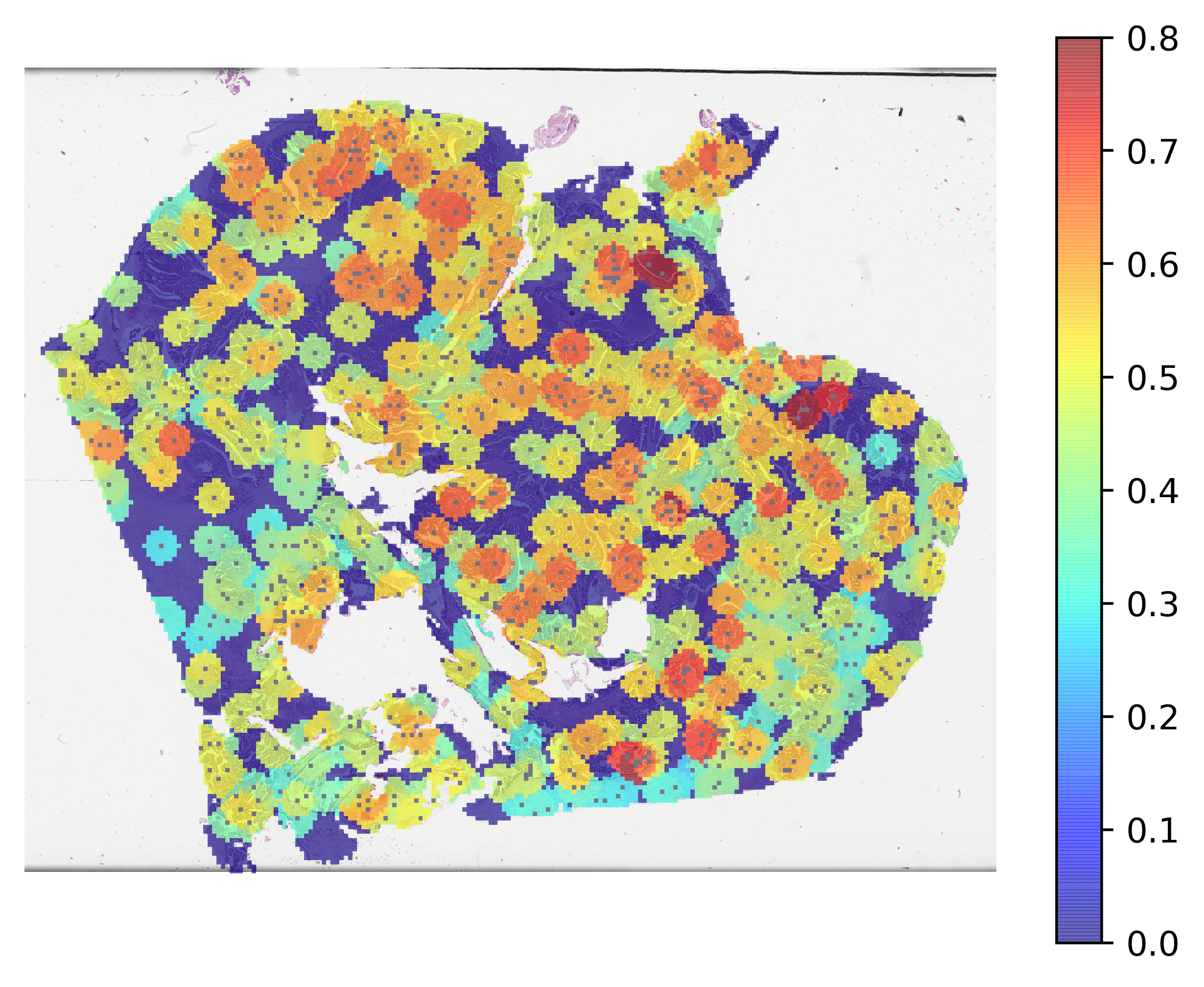}
\caption{800 Samples}
\label{fig:it12}
\end{subfigure}
\caption{Illustrative example of sampling weights generated through spatial sampling at different stages of the active sampling process. The closest 50 patches to each previous sample are assigned the corresponding sampling weight, and the proportion of random samples taken is 0.5, giving a relatively high level of exploration.}
\label{fig:sampling}
\end{figure}

\section{RESULTS}
\begin{table}[h]
\centering
\caption{\label{tab:metricresults} 3-fold cross-validated binary classification results using different evaluation approaches with the same baseline model (mean ± one standard deviation calculated using 100,000 epochs of bootstrapping from 50 repeats of the sampling approaches).}
\begin{tabular}{|c|c|c|c|c|}
\hline
 Method & AUC & Accuracy & Balanced Accuracy & F1 Score \\
\hline
\hline
Full MIL Evaluation & 0.8781 & 82.91\% & 80.08\% & 0.7472\\
\hline
DRAS-MIL & 0.8679±0.0035 & 81.63\%±0.64\% & 79.07\%±0.69\% & 0.7337±0.0093\\
\hline
Random Sampling & 0.8659±0.0034 & 81.63\%±0.62\% & 78.94\%±0.66\% & 0.7320±0.0089 \\
\hline
\end{tabular}
\end{table}

\begin{figure}[h]
\centering
\includegraphics[width=0.6\textwidth]{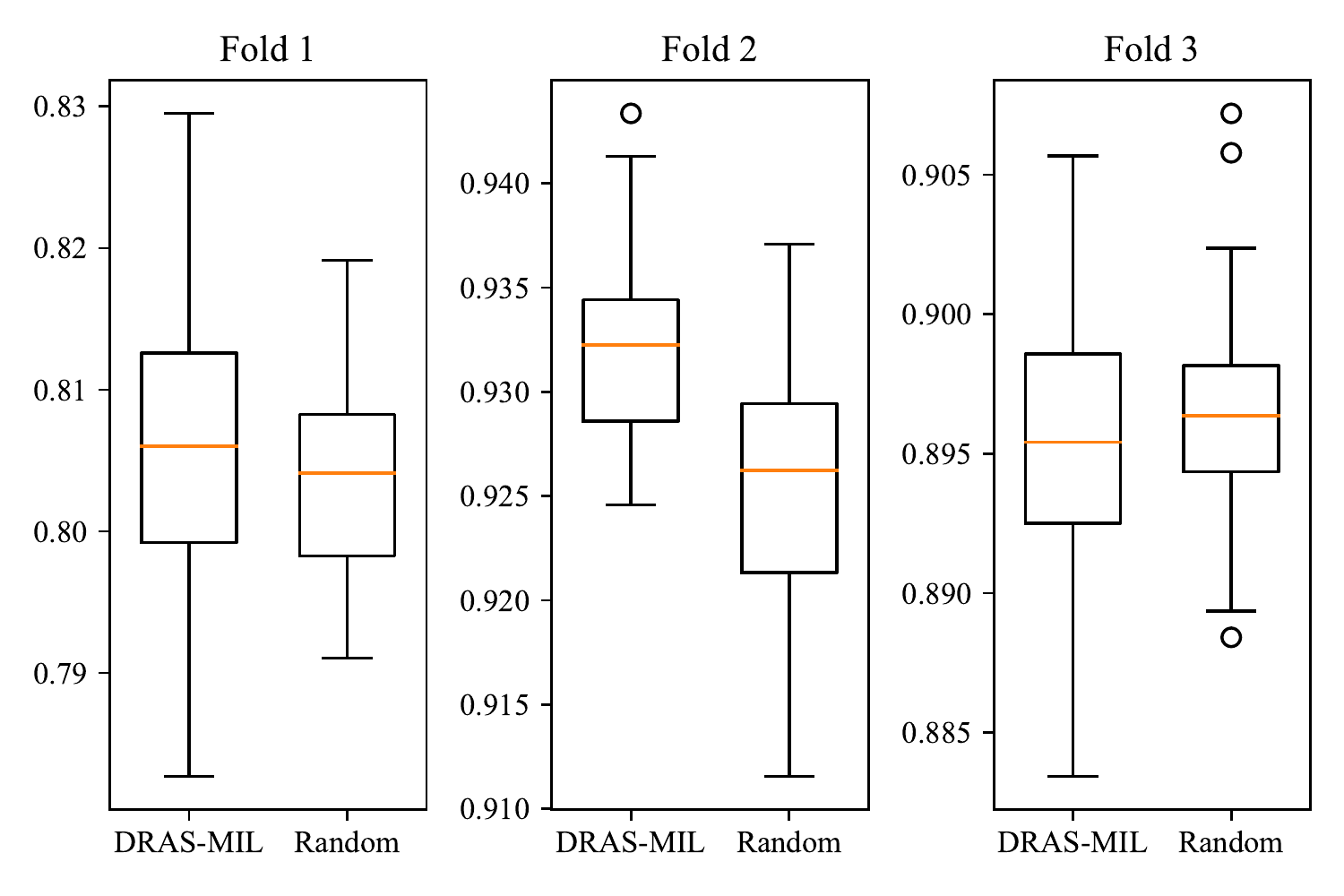}
\caption{Boxplot showing the distribution of AUC scores from 50 repeats of each sampling method during evaluation of the test set of each cross-validation fold.}
\label{fig:auc_results}
\end{figure}

\begin{figure}[h]
\centering
\includegraphics[width=\textwidth]{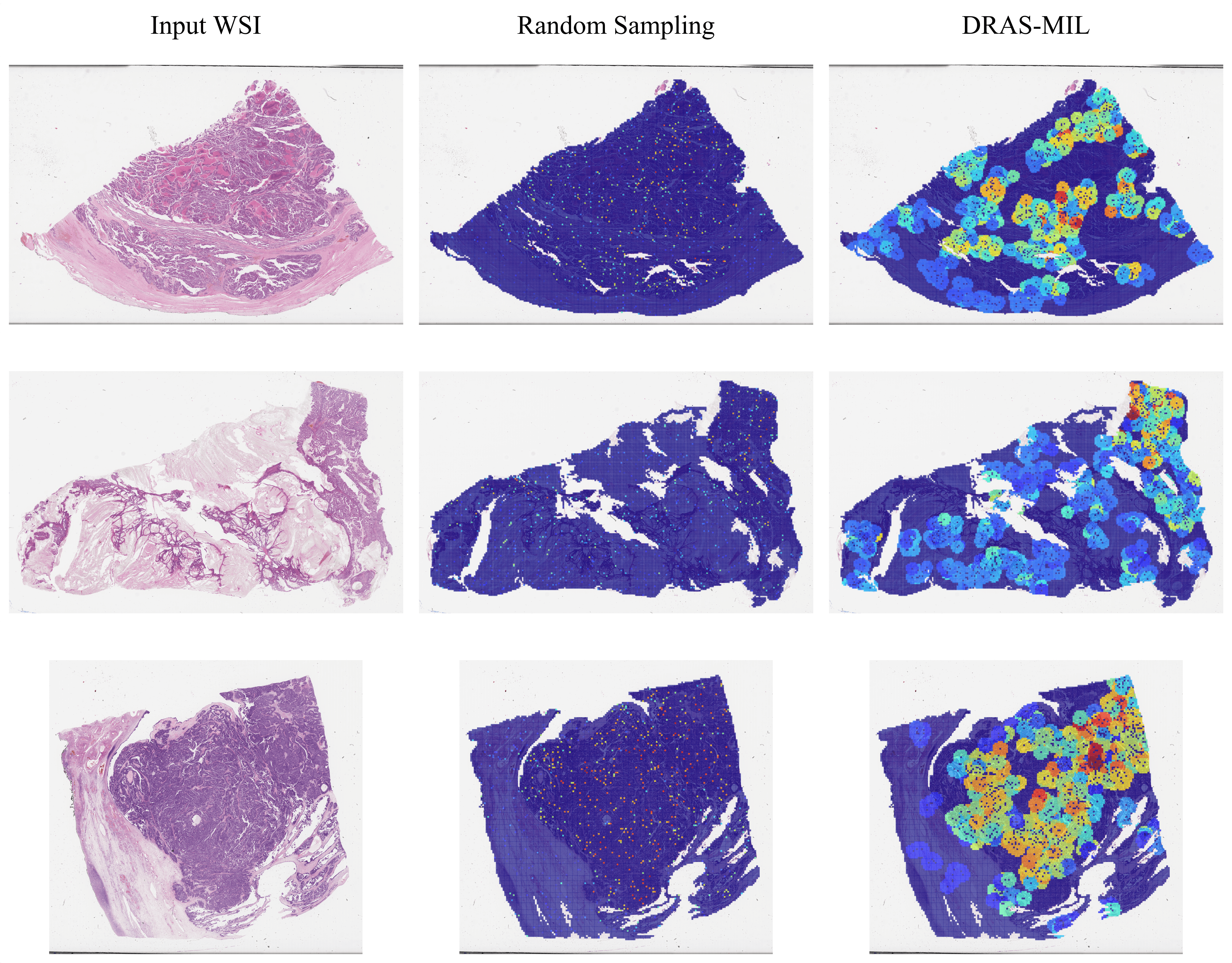}
\caption{Example WSIs with corresponding random sampling attention maps and DRAS-MIL sampling weight maps. Each method takes 800 samples per WSI. Reds indicate higher attentions/weights, and blues indicate lower attentions/weights.}
\label{fig:maps}
\end{figure}

We evaluated our method using four metrics - the area under the receiver operating characteristic curve (AUC), accuracy, balanced accuracy, and F1 score. We take AUC as our primary metric as this is a holistic classification metric which summarises true and false positive rates across all classification thresholds. Accuracy measures the proportion of correct predictions at a single threshold, with balanced accuracy being a weighted version which accounts for the different class sizes in the dataset. F1 score is a similar metric which takes the harmonic mean of precision and recall at a single threshold. Each of these metrics is a score between 0 and 1, meaning they can be expressed as percentages.  The cross-validated evaluation results are shown in Table \ref{tab:metricresults}. For each metric, DRAS-MIL performance was within 1.5\% of the whole slide processing performance. DRAS-MIL slightly outperformed random sampling for each metric, though was not consistently better across folds, as shown in the AUC distributions in Figure \ref{fig:auc_results}. Example sampling weight maps are shown in Figure \ref{fig:maps} and compared to corresponding random sampling attention scores. 

\begin{table}[h]
\centering
\caption{\label{tab:results} Efficiency experiment results, with each experiment run on the same set of 50 randomly selected WSIs. Each experiment was run three times and the median values were taken as the ground truth. All experiments use a GPU except where indicated otherwise. The minimum time and memory utilisation for each evaluation method on GPU is shown in bold.}
\begin{tabular}{|c|c|c|c|c|}
\hline
Evaluation Method &  Batch Size & \makecell{Total Evaluation Time \\ for 50 WSIs} & \makecell{Average Evaluation Time \\ per WSI} & \makecell{Maximum GPU \\ memory utilisation} \\
\hline
\hline
Default & 
\makecell{1 \\ 4 \\ 8 \\ 16 \\ 32 \\ 64} & \makecell{4h 37m \\ 2h 46m \\ 2h 33m \\ 2h 24m \\ \textbf{2h 20m} \\ 2h 21m} & \makecell{332.0s \\ 198.9s \\ 183.8s \\ 173.3s \\ \textbf{167.8s} \\ 169.3s} & \makecell{\textbf{340MB} \\ 342MB \\ 356MB \\ 471MB \\ 702MB \\ 1163MB}  \\
\hline
DRAS-MIL & \makecell{1 \\ 4 \\ 8 \\ 16 \\ 32 \\ 64} & \makecell{\textbf{47m} \\ 47m \\ 49m \\ 53m \\ 58m \\ 1h 1m} & \makecell{\textbf{56.0s} \\ 56.2s \\ 58.6s \\ 63.5s \\ 69.4s \\ 73.6s} & \makecell{\textbf{60MB} \\ 103MB \\ 161MB \\ 275MB \\ 506MB \\ 967MB} \\
\hline
Default (CPU) & 32 & 30h 32m & 2198.1s (36m 38s) & 0MB \\
\hline
DRAS-MIL (CPU) & 1 & 4h 8m & 298.0s (4m 58s) & 0MB \\
\hline
\end{tabular}
\end{table}

Table \ref{tab:results} shows the results of efficiency testing. On our computational benchmarking dataset of 50 WSIs, active sampling reduced GPU memory utilisation from a maximum of 340MB to 60MB. The best median run time for DRAS-MIL was 47 minutes for all 50 slides compared to 140 minutes for default MIL processing. This represents approximately 56s per WSI for active sampling and 168s per WSI for default MIL processing. The difference in run times is much greater when evaluating only using a CPU, with the total times of 4h 8m for active sampling and 30h 32m for full MIL evaluation, representing approximately 5 minutes and 37 minutes per slide, respectively.  It is unclear why increasing the batch size does not always result in faster evaluation times, but it may indicate an input/output bottleneck or issues with multithreading, and future work will investigate this phenomenon. Overall, GPU memory requirements are reduced by at least 82\% and evaluation time is reduced by 67\% when using a GPU, and by 86\% when using a CPU alone.

\section{DISCUSSION}
Our results indicate that DRAS-MIL can drastically reduce computational requirements for whole slide evaluation with minimal impact on classification accuracy. The baseline binary classifier achieved an accuracy of 83\%, which is favourable to the accuracy of two individual gynaecological pathologists reported in a previous study (78\%-79\%)\cite{Gavrielides2021}, but is surpassed by the 92\% external validation accuracy presented in a recent study \cite{Farahani2022}, where a much larger training dataset including manual annotation of tumour regions was employed. However, as all of these studies have used different datasets it is still not clear exactly how well these methods compare to each other and to the pathologists. DRAS-MIL achieved similar performance to the baseline classifier, with a reduction of 1.0\%-1.4\% in each of the four metrics evaluated. 

We found that completely random sampling also retains classification performance, with DRAS-MIL only marginally improving on average over random sampling for each metric. While random sampling appears to be a viable approach, DRAS-MIL has the advantage that its sampling maps improve explainability, as shown in Figure \ref{fig:sampling}. The relatively high classification performance of random sampling and DRAS-MIL may be influenced by our manual selection of slides to include high proportions of tumour tissue, making it likely that tumour tissue will be sampled by chance. Further investigations using less curated slides would be useful to better understand this process. 

In this research, we focused on maximising efficiency whilst minimising the resulting loss of accuracy. Sampling during evaluation can be useful in clinical practice, where advanced hardware is unavailable, by training a model externally and then running slide evaluation on standard clinical hardware. It is important to focus on reducing barriers to clinical implementation of computer-aided diagnostic tools as the underlying models are increasingly being shown to work at an expert level of accuracy in research settings, but are infeasible to implement in clinical settings. 
For this study, we were able to source slides from a limited quantity of patients, with 48 slides from only 11 patients for the least common subtype included. All slides were sourced from the same hospital and scanned with the same scanner, ensuring consistent laboratory H\&E staining protocols and minimising visual variability. However, the inclusion of non-adnexal tissue is a potential source of bias, as the model may classify based on the varying stromal tissue rather than the relevant tumour tissue. 

The subtyping model we have developed could be used in clinical practice as a second opinion for pathologists, especially those who are not gynaecological experts, to help distinguish the most common subtype from less common subtypes. In the case that an uncommon subtype is found, the advice of a gynaecological pathologist could then be taken, with the model having reduced the number of cases which needed to be referred to an expert. Before implementing this model we need to improve accuracy to better match the state-of-the-art methods, and ensure the model is robust to the visual variability seen in digital pathology. We also want to classify each subtype separately and better account for mixed subtypes, rare epithelial subtypes, and non-epithelial subtypes.

To meet our goals in future research we will increase the size of our ovarian carcinoma digital pathology dataset, allowing us to perform a rigorous analysis of five-class subtype classification. We will also validate on external data to understand the robustness of the models and methods. While our current research is focused on DRAS-MIL as an approach to make slide evaluation more efficient, our future work will also include using DRAS-MIL during model training to improve slide classification accuracy by training the model to better focus on discriminative regions, with an element of stochasticity which may help to reduce overfitting and mitigate for the effects of confounders and/or biases present in the data.



\section{CONCLUSION}
In this paper, we proposed an efficient slide evaluation approach, Discriminative Region Active Sampling for Multiple Instance Learning (DRAS-MIL). We demonstrated that it substantially increases the speed and reduces the computational requirements of evaluation of histopathology slides, without notably compromising ovarian cancer subtype classification performance. We found that completely random sampling gave similar performance, with just a slight decrease in AUC compared to DRAS-MIL, though random sampling is much less explainable than DRAS-MIL. We focused on efficiency as we believe that current slide classification methods are too slow and require too much computing power to be practicable in a clinical setting, especially when a GPU is not available. We hope that improving the efficiency of slide evaluation will aid the clinical deployment of computer-assisted diagnostic tools and reduce the associated costs.

\appendix 

\section{HYPERPARAMETER TUNING}
\label{sec:hyperparams}

To create our baseline classifier we performed 500 experiments of random hyperparameter tuning to minimise the loss on the validation set of the first cross-validation fold. The tuned hyperparameters for model training were learning rate, regularisation rate, and drop out, with sampling distributions shown in Table \ref{table:tuning}. This tuning was completed twice, once using a cross-entropy loss and once using a \emph{balanced} cross-entropy loss. The best resulting model used a cross-entropy loss, with a learning rate of 0.0038, regularisation rate of 0.00079, and drop out of 0.020. This model achieved an overall AUC of 0.902 on the validation set and 0.878 on the test set. 

To optimise hyperparameters for DRAS-MIL we performed 200 experiments of random hyperparameter tuning to maximise validation AUC. The hyperparameters included were the number of resampling iterations, the number of spatial neighbours to which sampling weights were applied, the proportion of samples which were taken at random, and the reduction in this proportion per iteration, as shown in Table \ref{table:tuning}. To help account for the randomness inherent to active sampling, each tuning experiment was repeated 30 times and the average AUC was taken. The best model used 10 resampling iterations, 64 sampling neighbours, 0.29 random sampling rate and 0.36 random sampling delta.  

\begin{table}[h]
\centering
 \caption{Hyperparameters tuned using a random search. The first three were tuned during baseline model training, and the subsequent five were tuned during slide evaluation. *Samples per iteration was not tuned independently, but calculated based on the number of resampling iterations to always give 640 samples, with an extra 160 samples taken in the final step of the evaluation.}
 \label{table:tuning}
 \begin{tabular}{|c|c|c|c|} 
 \hline
\textbf{Category} & \textbf{Hyperparameter} & \textbf{Function} & \textbf{Distribution} \\
 \hline
 \hline
  & Learning Rate & \makecell{The rate of change \\ for model parameters} & Log-Uniform (1e-5, 1e-2) \\
  \cline{2-4}
  Training & Regularisation Rate & \makecell{The level of weight \\ decay in Adam optimizer} & Log-Uniform (1e-10, 1e-2)  \\
  \cline{2-4}
  & Drop Out & \makecell{The proportion of weights to \\ drop to reduce overfitting} & Uniform (0.00, 0.99)  \\
  \hline
  & Resampling Iterations & \makecell{The number of \\ resampling iterations} & Choice [2,4,6,8,10,12,16]  \\
  \cline{2-4}
  & \makecell{Samples per \\ Iteration*} & \makecell{The number of \\ samples per iteration} & Choice [320,160,107,80,64,53,40] \\
  \cline{2-4}
  Sampling & Sampling Neighbours & \makecell{The number of nearest \\ neighbours assigned weights} & Choice [4,8,16,32,48,64]  \\
  \cline{2-4}
  & Sampling Random & \makecell{The proportion of samples \\ which are randomly sampled} & Uniform (0.00, 0.75)  \\
  \cline{2-4}
  & Sampling Random Delta & \makecell{The reduction in Sampling \\ Random each iteration} & Log-Uniform (0.0001, 0.5)  \\
 \hline
\end{tabular}
\end{table}

\section*{CODE AVAILABILITY}
All of our code is available at \url{www.github.com/scjjb/DRAS-MIL} and can be easily adjusted to classify other types of histopathology slides. This implementation builds on the code provided at \url{www.github.com/mahmoodlab/CLAM} \cite{Lu2021}. 

\section*{AUTHOR CONTRIBUTIONS}
JB and NR conceived the AI approach. JB wrote code and performed experiments. KA collected histopathological data and clinical metadata. KA and NO reviewed morphological diagnoses and provided clinical guidance. KZ and GH provided oncological support. JB wrote the manuscript with feedback and contributions from all other authors.  

\section*{ACKNOWLEDGEMENTS}
This research was funded in part by the UKRI Engineering and Physical Sciences Research Council (EPSRC) [EP/S024336/1]. For the purpose of open access, the author has applied a Creative Commons Attribution (CC BY) licence to any Author Accepted Manuscript version arising from this submission. Data collection was funded by the Tony Bramall Charitable Trust. The authors are grateful to Mr Mike Hale and Professor Darren Treanor for the provision of slide scanning support. This work uses data provided by patients and collected by the NHS as part of their care. The use of data for this project received approval from the Wales Research Ethics Committee (REC; reference 18/WA/0222) and secured Confidentiality Advisory Group approval (CAG; reference 18/CAG/0124).

\bibliography{main} 
\bibliographystyle{spiebib} 

\end{document}